\documentclass[11pt]{article}
\usepackage{moriond,epsfig}
\usepackage{mystyle}

\def\Journal#1#2#3#4{{#1} {\bf #2}, #3 (#4)}

\def\NIMA{{\em Nucl. Instrum. Methods} A}

\def\PLB{{\em Phys. Lett.}  B}

\def\PRD{{\em Phys. Rev.} D}
\def\ZPC{{\em Z. Phys.} C}
\def\EURJ{{\em Eur. Phys. J.} C}
\def\JHEP{{\em JHEP}}
\def\MPLA{{\em Mod. Phys. Lett.} A}

\def\be{\begin{equation}}
\def\ee{\end{equation}}
\def\bea{\begin{eqnarray}}
\def\eea{\end{eqnarray}}

\def\scr{\scriptstyle}
\def\sscr{\scriptscriptstyle}
\def\babar{\ifmath{BABAR}}
\def\pipiu{\ifmath{\pi^{\scr +}}}
\def\pimeno{\ifmath{\pi^{\scr -}}}

\def\epiu{\ifmath{e^{\scr +}}}
\def\emeno{\ifmath{e^{\scr -}}}
\def\mupiu{\ifmath{\mu^{\scr +}}}
\def\mumeno{\ifmath{\mu^{\scr -}}}
\def\kpiu{\ifmath{K^{\scr +}}}
\def\kmeno{\ifmath{K^{\scr -}}}

\def\eetomumugamma{\ensuremath{\epiu\emeno\rightarrow\mupiu\mumeno\gamma}}
\def\fourpi{\ensuremath{2\pipiu 2\pimeno}}
\def\eetopioni{\ensuremath{\epiu\emeno\rightarrow\fourpi}}
\def\eetofourhad{\ensuremath{\epiu\emeno\rightarrow 2h^{\scr +}2h^{\scr -}}}
\def\fourK{\ensuremath{2\kpiu 2\kmeno}}

\def\twopitwoK{\ensuremath{\kpiu\kmeno\pipiu\pimeno}}

\def\zetazero{\ensuremath{\Z^{\scr 0}}}
\def\upsifours{\ifmath{\Upsilon(4S)}}
\def\jpsi{\ifmath{J/\psi}}
\def\eetojpsi{\ensuremath{\epiu\emeno\rightarrow \jpsi}}
\def\ERRE{\ensuremath{{\mathrm R} ~=~ \sigma(\epiu\emeno\to{\mathrm hadrons}) / \sigma(\epiu\emeno\to\mupiu\mumeno)}}
\def\gmenodue{\ensuremath{ (g-2)_{\scr \mu}}}
\def\alphaqed{\ensuremath{ \alpha(M_{\scr Z}^{\scr 2})}}
 
\begin{document}
\vspace*{4cm}
\title{INTIAL STATE RADIATION AND INCLUSIVE HADRON PRODUCTION MEASUREMENTS AT BABAR.}

\author{ F. ANULLI }

\address{University of Perugia and
 Laboratori Nazionali di Frascati dell'INFN,\\ 
 via E.Fermi 40, I-00044 Frascati (Rm), Italy }

\maketitle\abstracts{
The status of analysis of processes with hard photon emitted from the initial state (ISR) at \babar\ 
is presented. We tag events by the presence of a hard photon in the detector, then reconstruct 
 $\mupiu \mumeno$ and several exckusive hadronic final states.
The invariant mass of the final state determines an effective collision center of mass energy 
at which these measurements can be compared with results from direct $\epiu\emeno$ annihilation process. 
The first results on \eetofourhad, where $h=\pi,K$ and on \eetojpsi, 
obtained with a sample of 89.3 \ifb, are presented.\\ 
Measurements of inclusive $\eta,~\pipiu, ~\kpiu ~{\mathrm {and}} ~p/\bar{p}$ production 
cross sections below the \upsifours\ resonance are also presented.
These measurements have  nearly complete momentum coverage and 
precision comparable to the best measurements at higher energies.
They therefore allow for precise tests of QCD predictions and  
fragmentation models at 10.54 GeV and of their scaling properties.}

\section{Initial State Radiation processes at \babar.}\label{sec:ISR}
Initial state radiation (ISR) processes can be effectively used to measure \epiu\emeno\ annihilation
at high luminosity \epiu\emeno\ storage rings, 
such as the $B$-$factory$ PEP-II~\cite{bib:ISR1,bib:ISR2,bib:ISR3}.
A large mass range is accessible in a single experiment, contrary to the case with fixed energy colliders,
which are optimized only in a limited region. In addition, the broad-band coverage may result also in
greater control of systematic effects because only one experimental setup is involved.\\
The ISR physics program consists mainly on light hadron spectroscopy 
and measurement of the ratio \ERRE, which provides the experimental input to dispersion 
integrals for computation of the hadronic contribution to the theoretical estimation of  
the muon magnetic moment anomaly,  $a_{\scr \mu} = \gmenodue/2$ 
and of the running of the electromagnetic coupling to the $Z$ pole, \alphaqed.\\ 
The ISR cross section for a particular final state $f$ is related to the cross section 
for the direct annihilation $\epiu\emeno\to f$  through   
\begin{equation}
\frac{d\sigma(s,x)}{dx} ~=~ W(s,x) ~\sigma_f(s^{\scr \prime}),  ~~~~~s^{\scr \prime}=s(1-x); 
\label{eq:ISRcs}
\end{equation}
where $x = 2 E_{\scr \gamma}^{\star}/\sqrt{s}$, $ E_{\scr \gamma}^{\star}$ is the
 radiated photon energy in the 
nominal center-of-mass frame and $\sqrt{s}$ the nominal c.m. energy of the collider. 
The quantity  $s^{\scr \prime}=s(1-x)$
represents the mass squared of the final state system, $f$. The radiator function $W(s,x)$ describes 
the virtual photon energy spectrum and can be computed to an accurcy better than 1\%.
The direction of radiated photon is peaked along the initial beams, 
but for $\sqrt{s} \simeq 10 ~{\mathrm{GeV}}$ 
the fraction at large angle is  relatively large. 
It has been shown a 10-15\% acceptance for these photons in \babar.\\
The measurement of the corresponding leptonic process \eetomumugamma\ provides the ISR luminosity. 
Thus, the Born cross section for a hadronic final state $\sigma_{\scr f}(s^{\scr \prime})$ is given by
\begin{equation}
\sigma_f(s^{\scr \prime}) ~=~ 
\frac{\Delta N_{\scr f\gamma} ~\epsilon_{\scr \mu\mu} ~(1+\delta_{\scr FSR}^{\scr \mu\mu})}{\Delta N_{\scr \mu\mu\gamma} 
~\epsilon_{\scr f} ~(1+\delta_{\scr FSR}^{\scr f})} ~~ \sigma_{\mu\mu}(s^{\scr \prime})
\label{eq:ISRfsCrossSec}
\end{equation}
where $\Delta N_{\scr f\gamma}$ ($\Delta N_{\scr \mu\mu\gamma}$) are the number of detected 
$f\gamma$ ($\mu\mu\gamma$) events 
in the bin of width $\Delta s^{\scr \prime}$ centered at $s^{\scr \prime}$. 
$\epsilon_{\scr f}$ ($\epsilon_{\scr \mu\mu}$)
and $\delta_{\scr FSR}^{\scr f}$ ($\delta_{\scr FSR}^{\scr \mu\mu}$) are respectively the  
detection efficiencies and
fractions of events when the hard photon is emitted by final-state particles. 
The latter quantity can be sizable for the $\mu\mu$ channel, but negligible for most 
of the low energy hadronic states,
 which have vanishingly small cross sections at the nominal machine energy.\\

\begin{figure}
\begin{minipage}[b]{0.48\textwidth}
\vspace{-0.6cm}
\includegraphics[height=.28\textheight]{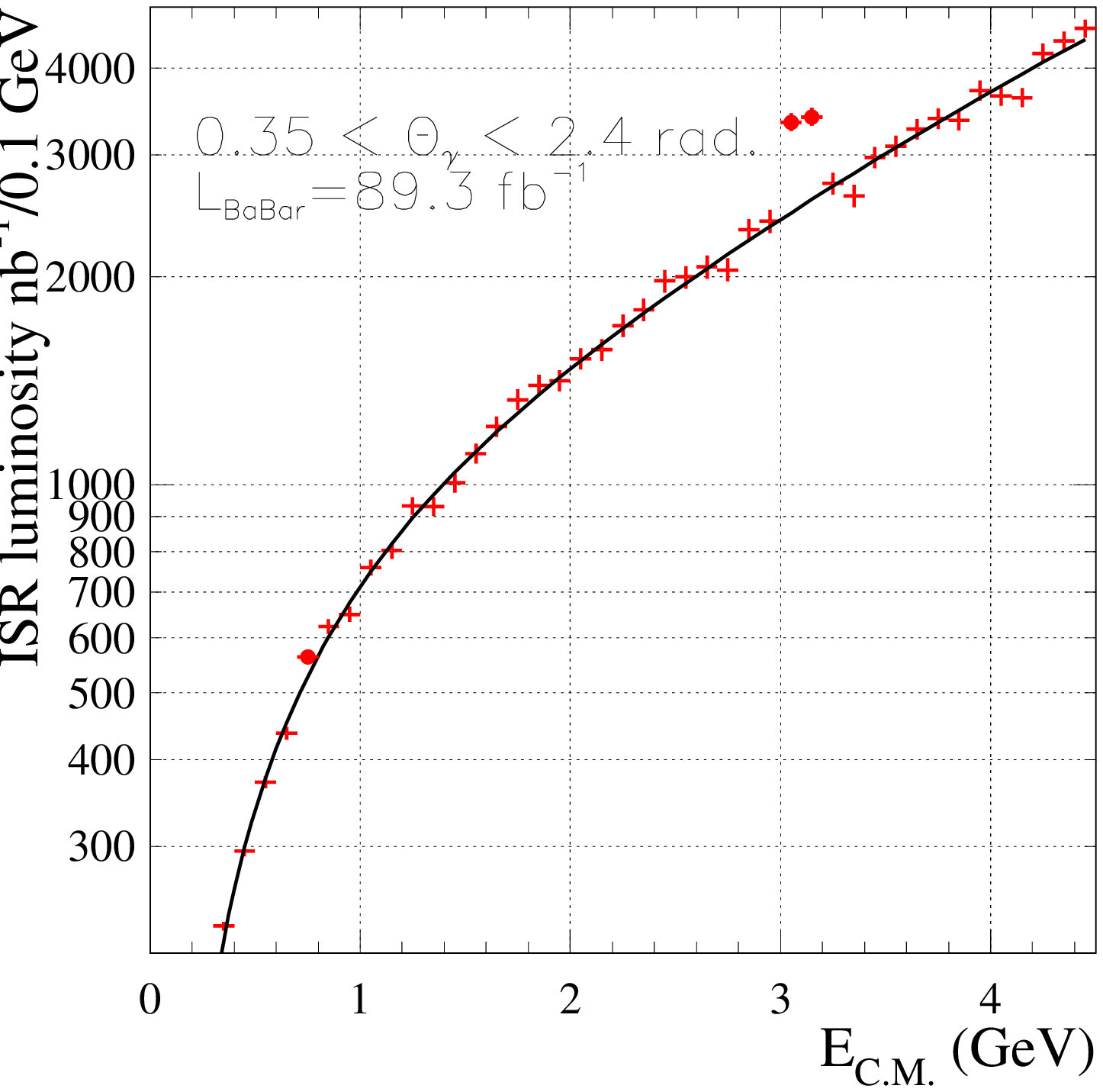}
\end{minipage}
\hfill
\begin{minipage}[b]{0.48\textwidth}
\includegraphics[height=.28\textheight, width=1.0\textwidth]{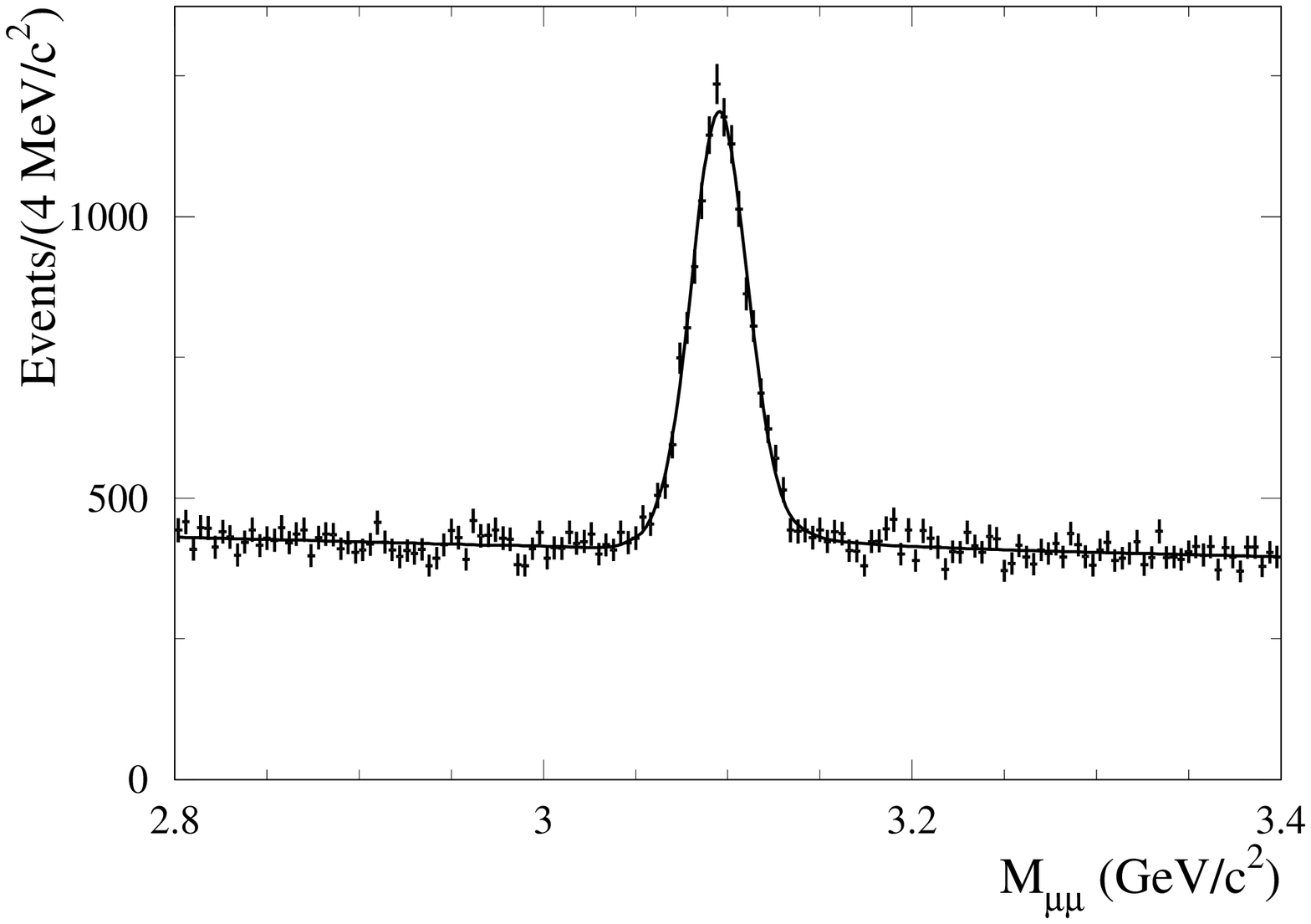}
\end{minipage}
\caption{Left: measured ISR luminosity, integrated over 0.1 GeV for 89.3 \ifb\ of \babar\ integrated
luminosity. On the right: the reconstructed \mupiu\mumeno\ invariant mass distribution in the \jpsi\ region.}
\label{fig:ISRlumi}
\end{figure}

\subsection{The di-muon final state and ISR luminosity at \babar}
The data used in the analysis presented here correspond  to an integrated luminosity of 89.3\ifb\, collected 
 both at the \upsifours\ and in the nearby continuum, with the \babar\ detector 
at the PEP-II asymmetric $B$-factory.\\
The \babar\ detector is described elsewhere~\cite{bib:BABAR}. The information from the tracking system
(Silicon Vertex Tracker and Drift Chamber) is used to measure angles and momenta of charged particles.
The quartz Cherenkov radiator (DIRC) is the main subsystem for particle identification 
($K/\pi$ separation is an essential ingredient for these studies).  
Muons are identified in the Resistive Plate Counters
installed in the magnet yoke of the \babar\ solenoid, 
while photons are detected in the CsI calorimeter.\\
Events are tagged by detecting a photon with an energy in the c.m. system larger than 3 GeV. 
The invariant mass of the dimuon system determines the effective collision energy. 
A disadvantage deriving from the use of ISR events is the invaraiant mass resolution, comparing to the
case of fixed energy machines. The resolution can  be improved by a kinematics fit constraining 
the mass recoiling against the reconstructed final state to be equal to zero, that is the photon mass.\\
After the kinematics fit the invariant mass resolution improves from $16~{\mathrm{MeV}}/c^{\scr 2}$ to 
$8~{\mathrm{MeV}}/c^{\scr 2}$ at the \jpsi\ mass. In addition the fit effectively removes background from 
$\tau^{\scr +} \tau^{\scr -}$ events.
Fig.~\ref{fig:ISRlumi}(left) show the derived spectrum of ISR luminosity per 100 MeV energy bins. The \babar\ 
luminosity of 89 \ifb\ is equivalent  to an \epiu\emeno\ machine energy scan in steps of 0.1 GeV, with 
a luminosity  integral per point varying from 0.7 \ipb\ at 1 GeV up to 3.6 \ipb\ at 4 GeV. 
This luminosity integral provides already a statistically competitive hadron sample, especially in the 
energy range between 1.4 and 3.5 GeV, where very few  data are presently available. The systematic error is
estimated to be of the order of 3\%(5\% for energies below 1 GeV).\\ 
  
\subsection{Hadronic channels}
A very rich program can be exploited in \babar\ from the study of hadronic final states:
spectroscopy, form factor measurements, search for exotic states, etc. 
Besides, measurements of exclusive hadronic channels constitute the main approach for measuring $R$.
Currently the major hadronic final states are under study 
($\pi\pi, KK, 4\pi, 5\pi, 6\pi, 2K2\pi, 4K, p\bar{p}, KK\pi$) and more are planned.\\
Preliminary results are available for the final states with four charged hadrons, namely  
\fourpi, \fourK, \twopitwoK. The discrimination between the three final states is done 
on the basis of the particle identification and on the kinematics fit results 
for the different mass hypothesis.\\
\begin{figure}
\begin{minipage}[b]{0.48\textwidth}
\vspace{-0.6cm}
\includegraphics[height=.28\textheight]{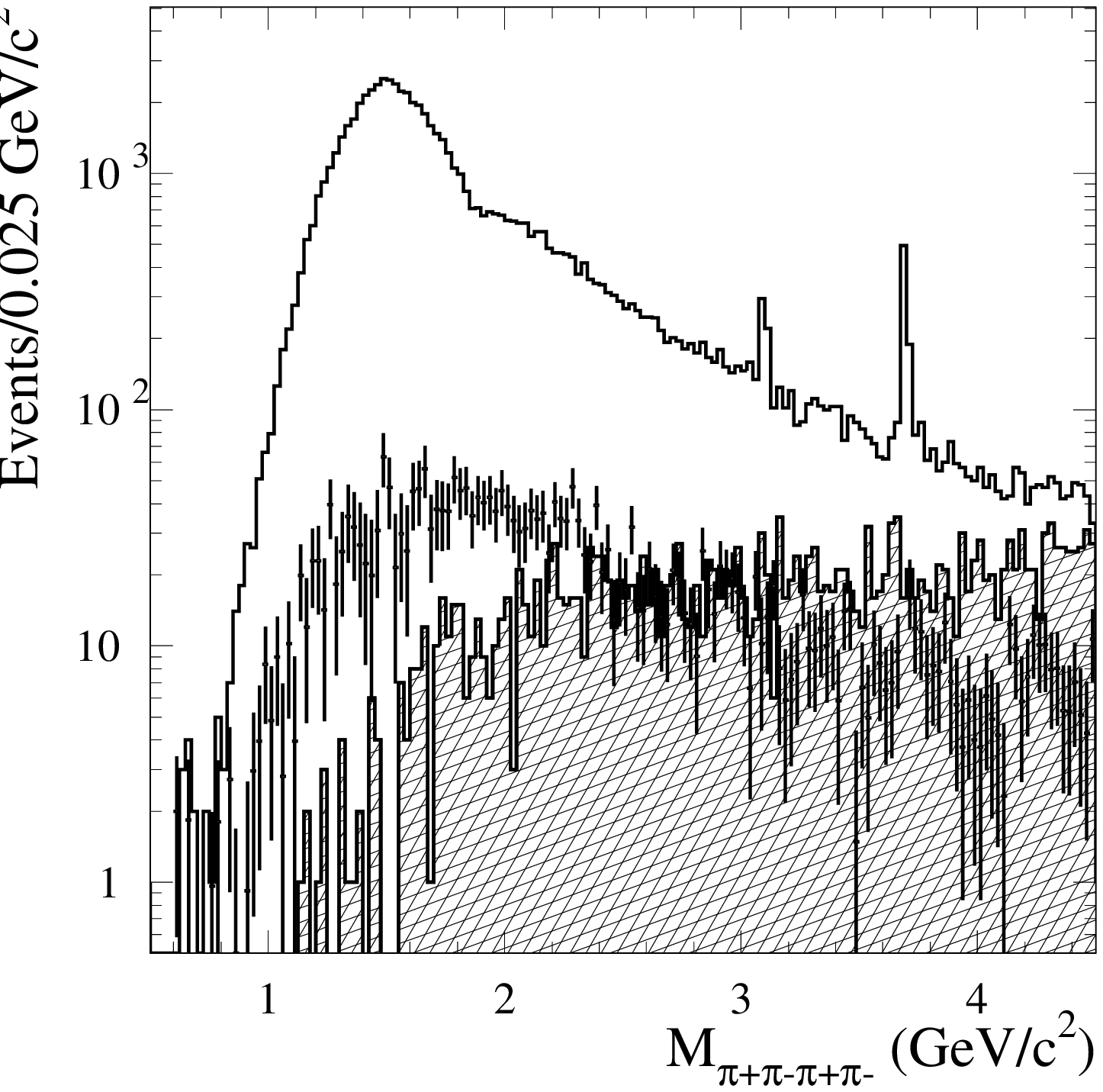}
\end{minipage}
\hfill
\begin{minipage}[b]{0.48\textwidth}
\includegraphics[height=.28\textheight]{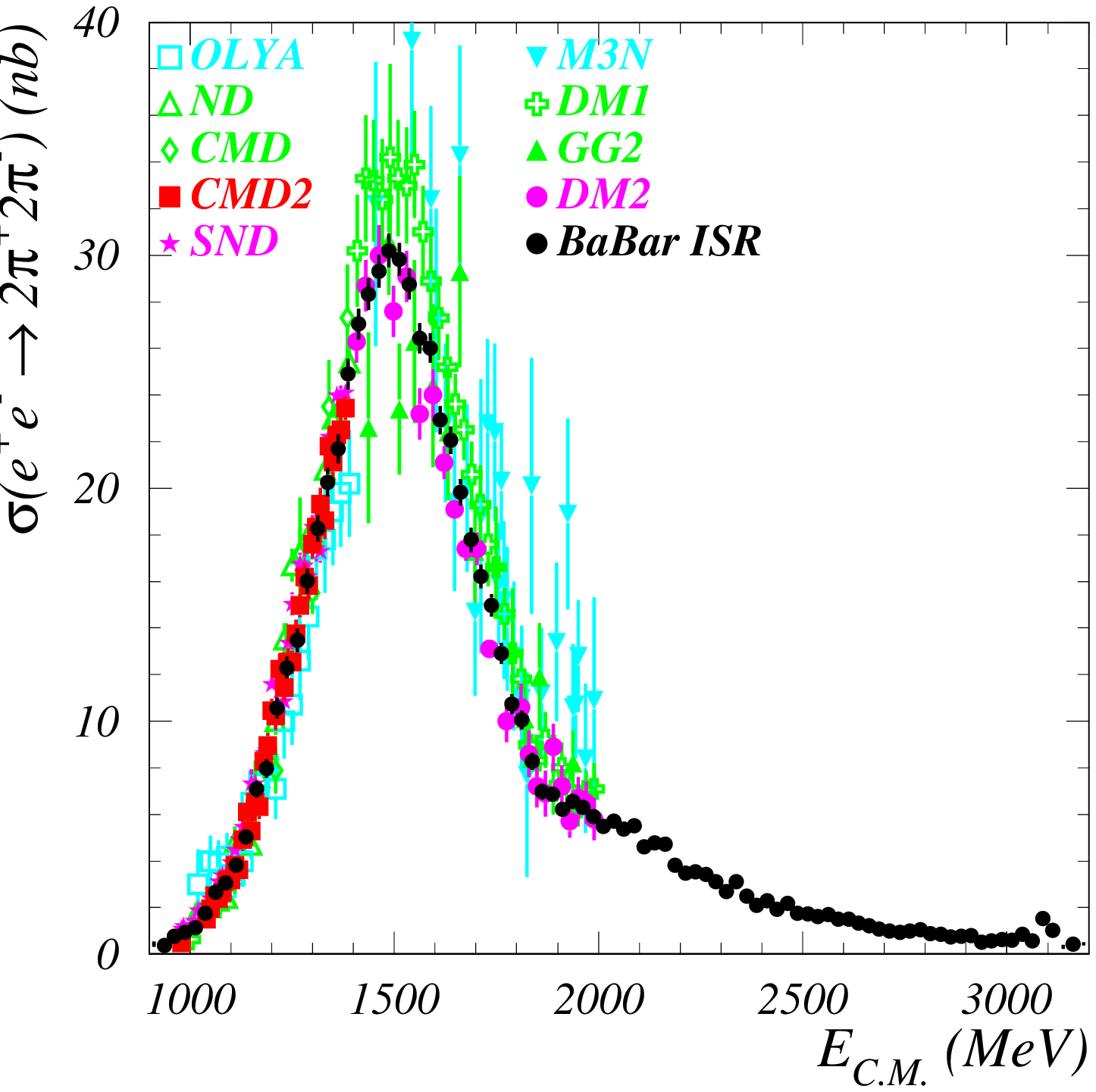}
\end{minipage}
\caption{Left: the four-pion invariant mass distribution. The points indicate the estiimated ISR-type backgrund, 
while the cross-hatched histogram corresponds to the non-ISR background. 
~ On the right: \eetopioni cross section as a function of energy obtained with \babar\ ISR data (black dots)
in comparison with all previous \epiu\emeno\ data.}
\label{fig:fourpisigma}
\end{figure}
No muon identification was applied since no corresponding background is expected in this sample.
The background from other hadronic channels is small and is subtracted in each mass bin.
Overall, more than 70000 events have been selected leading to small statistical uncertainties. 
The estimated systematic error is about 5\% in the energy region between 1 and 3 GeV, 
dominated by the luminosity determination.
Fig.~\ref{fig:fourpisigma}(right) shows the derived cross section for the 4 pions channel
in  25 MeV steps. \babar\ data are in good agreement with previous available results. 
Moreover, \babar\ is the only experiment which cover the entire energy range, with an accuracy
comparable to the latest precise 
results from CMD-2~\cite{bib:CMD2} and SND~\cite{bib:SND} below 1.4 GeV,
and much better accuracy than older results from DCI and ADONE above 1.4 GeV.\\
The hadronic contribution for this particular channel evaluated using all available \epiu\emeno\
data in 0.56-1.8 GeV energy range is~\cite{bib:davieretal}
$a_{\scr \mu}^{\scr hadr} = (14.21\pm 0.87_{\scr exp}\pm 0.23{\scr rad}) 10^{\scr -10}$, while the $\tau$
data give $a_{\scr \mu}^{\scr hadr} = (12.35\pm 0.96_{\scr exp}\pm 0.40_{\scr SU2}) 10^{\scr -10}$. 
The \babar\ data in the same energy region give instead~\cite{bib:davier4pi}
$a_{\scr \mu}^{\scr hadr} = (12.95\pm 0.64_{\scr exp}\pm 0.13{\scr rad}) 10^{\scr -10}$, 
leading to a substantial improvement.\\

\subsection{Measurement of \jpsi\ width}

For a narrow state of mass M such as \jpsi, decaying into a final state $f$, the Breit-Wigner can be approximated
by a $\delta$ function and the production cross section can be written as:
\begin{equation} 
 \sigma_{\scr \epiu\emeno\to\jpsi\gamma\to f\gamma} 
              ~=~ \frac{12\pi^{\scr 2}\Gamma_{\scr ee}B_{\scr f}}{M s} ~W(s,x_{\scr 0});
\label{eq:sigmajpsi}
\end{equation}
where $x_{\scr 0} = 1 - M^{\scr 2}/s$. The ratio of \mupiu\mumeno\ events from \jpsi\ peak (about 7800 events observed 
in 89\ifb, fig.~\ref{fig:ISRlumi}(right)) to the continuum allows to extract the product~\cite{bib:jpsimumu}
\begin{equation} 
\Gamma(\jpsi\rightarrow\epiu\emeno) ~ B_{\scr \jpsi\rightarrow\mupiu\mumeno} 
~=~ (0.330\pm0.008_{\scr stat} \pm 0.007_{\scr syst}) {\mathrm{keV}}.
\label{eq:GammaBF}
\end{equation}
Background estimation, \jpsi\ line shape, radiative corrections and Monte Carlo statistics constitute the main 
contributions to the systematic error. Using the world average values for the branching fraction
$B_{\scr \jpsi\rightarrow\mupiu\mumeno}$
and $B_{\scr \jpsi\rightarrow\epiu\emeno}$, we derive the \jpsi\ electronic and total widths,
$\Gamma_{\scr \jpsi\rightarrow\epiu\emeno} = (5.61\pm 0.20) ~{\mathrm{keV}}$ 
and $\Gamma_{\scr \jpsi} = (94.7\pm 4.4) ~{\mathrm{keV}}$, which represent an improvement with 
respect to the present world average values of respectively 
$(5.26\pm 0.37) ~{\mathrm{keV}}$ and $(87 \pm 5) ~{\mathrm{keV}}$.\\

\section{Inclusive hadron production studies}
\babar\ has performed measurements of inclusive production cross sections and fractions of etas and charged pions, 
kaons and protons, both from \upsifours\ decays and in continuum (at $E_{\scr CM} = 10.54$ GeV, 
below the $B$-$\bar{B}$ production threshold). 
The quark-antiquark colorless system created in the continuum events ``fragments'' into a 
number of primary hadrons, which then decay into stable hadrons. 
The fragmentation process has been extensively studied at higher energies,
in particular in the \zetazero\ region and a number of models have been developped and tested.\\
Models have been tuned for high energies, but fundamental tests, such as the study of 
scaling properties, need accurate
inclusive measurement of hadron  productions also at lower energies.
Previous measurements performed at Argus and Cleo at an energy of $\sim 10$ GeV were not able
to cover the full kinematics range. 
\babar\ data are consistent with those results and extend coverage up
to the highest momentum values with an accuracy at few percent level.\\
\begin{figure}[htb]
\begin{center}
\begin{tabular}{c c c}
\includegraphics[height=.2\textheight,width=.31\textwidth]{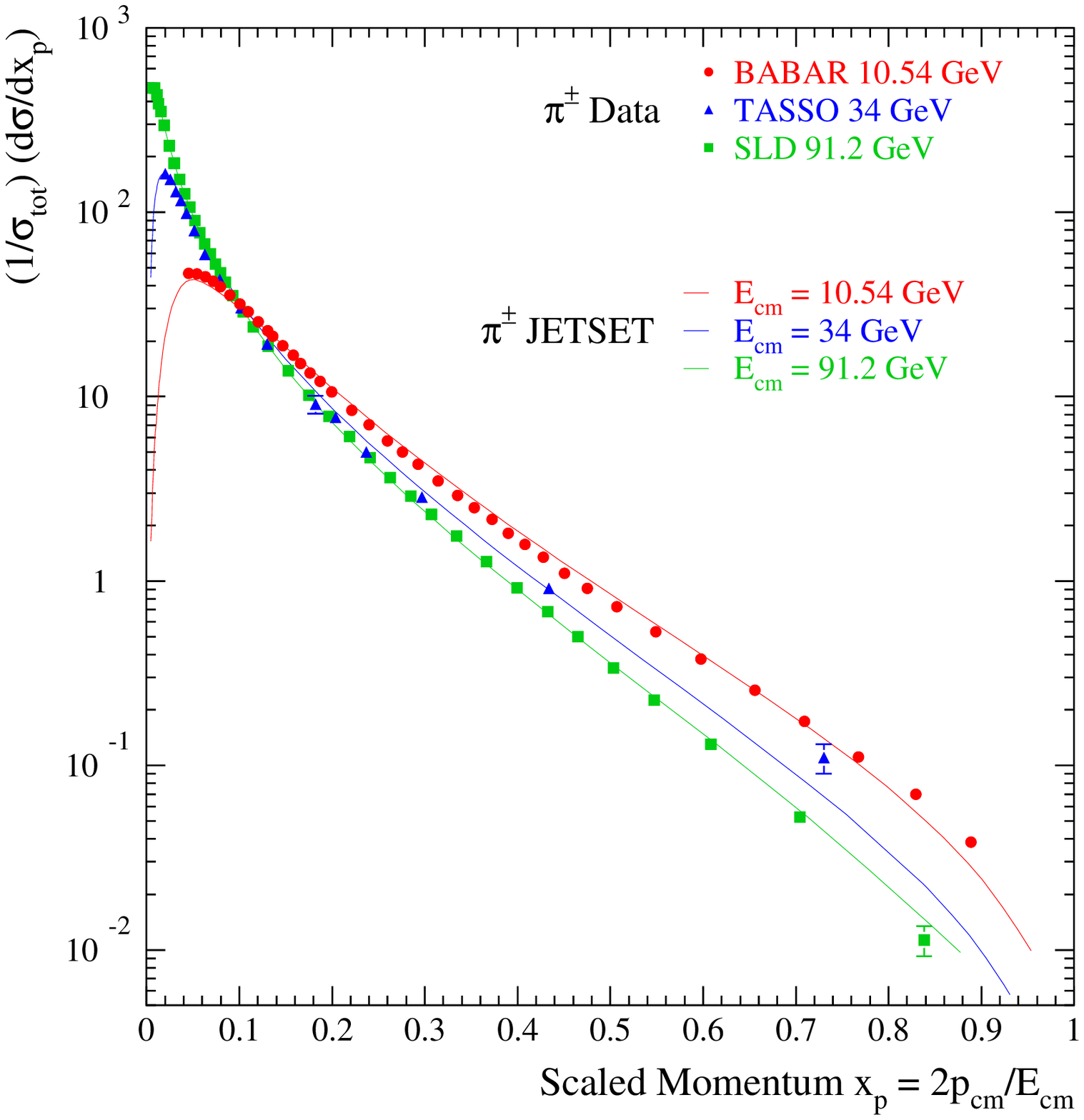} &
\includegraphics[height=.2\textheight,width=.31\textwidth]{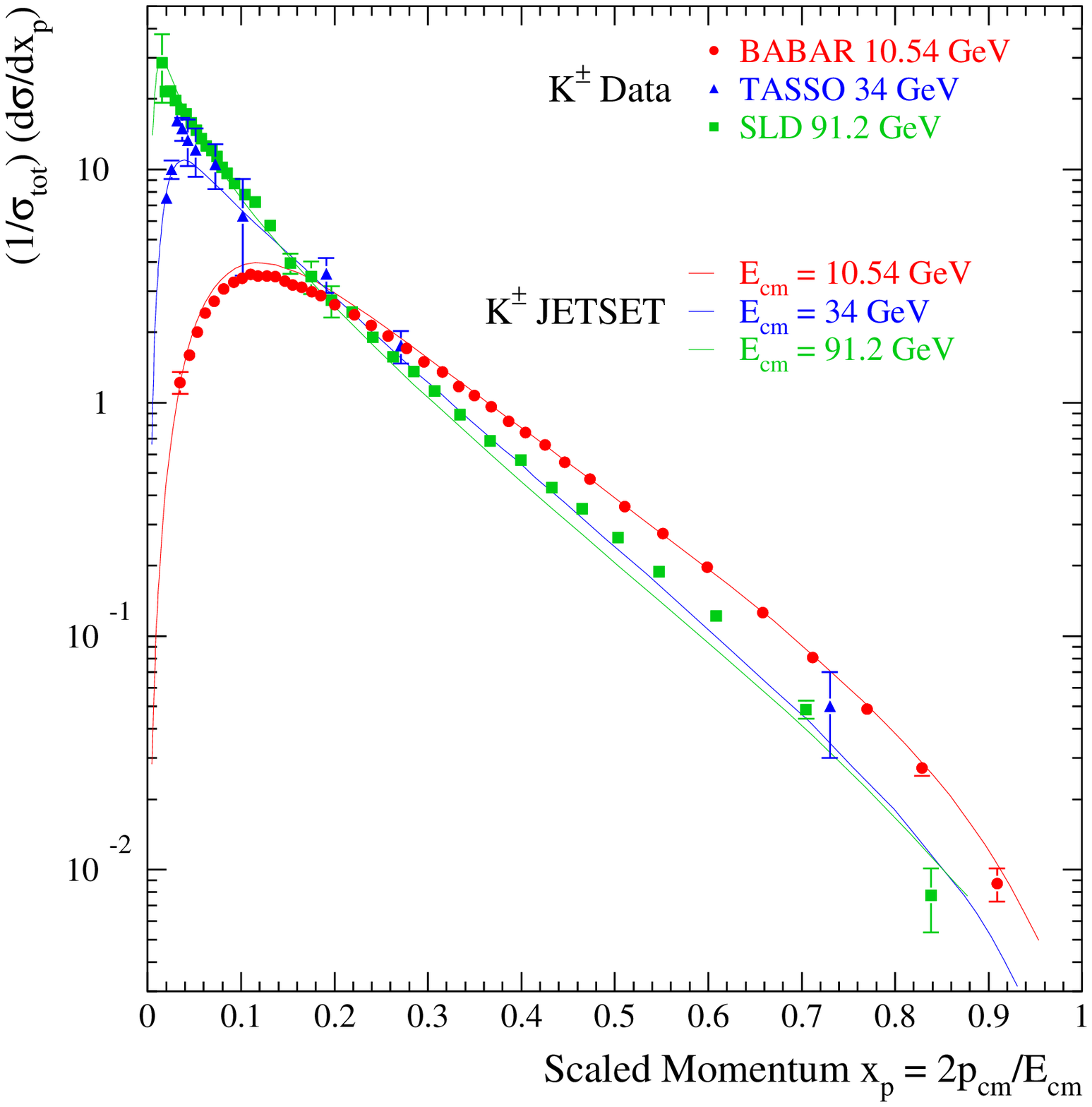} &
\includegraphics[height=.2\textheight,width=.31\textwidth]{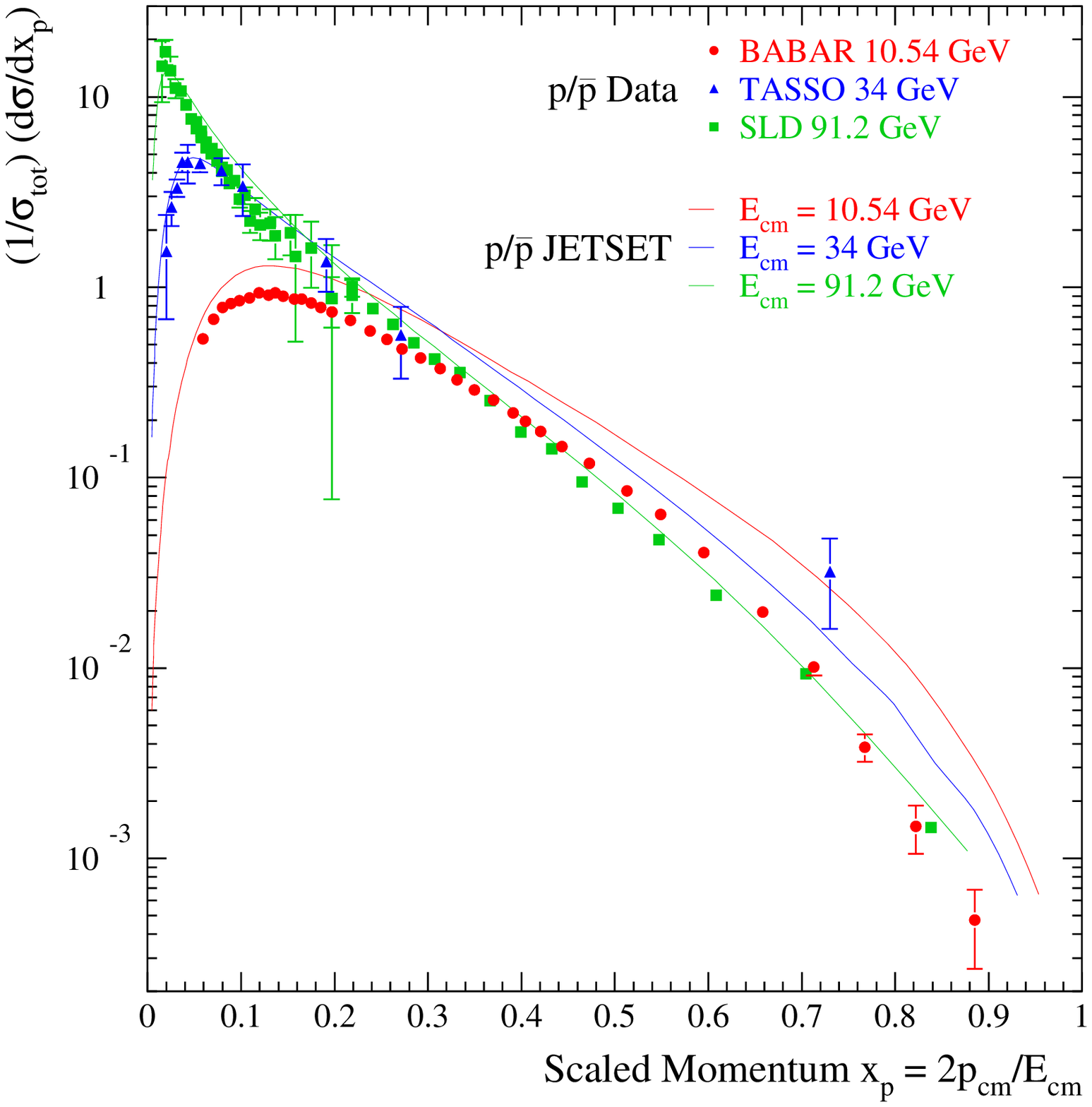} 
\end{tabular}
\end{center}
\caption{Differential pion (left), kaon (middle) and proton (right) 
cross section from continuum data in \babar\ (red points), compared with
results by TASSO at c.m. energy of 34 GeV (blu triangles) and by SLD at 91.2 GeV (green squares). 
Predictions by JETSET Monte Carlo for different energies are superimposed with corresponding color code.}
\label{fig:csvsxp}
\end{figure}
Fig.~\ref{fig:csvsxp} shows the differential cross sections in continuum events for 
$\pi, K ~{\rm and}~ p(\bar{p})$
as a function of $x_{\scr p}$, the momentum scaled by half of the center-of-mass energy.
\babar\ data, obatined from a sample of 0.9 \ifb, are shown along with those from 
experiment at higher energies, namely TASSO~\cite{bib:TASSO} at 34 GeV 
and  SLD~\cite{bib:SLD}  at 91.2 GeV. 
The nearly full  kinematics range coverage and the comparable precision between \babar\ and SLD data, 
allow to study scaling properties of hadronization and to test the predictions by the various
fragmentation models.
Hadronization should be scale invariant. Therefore, cross sections are expected to differ only 
because of small scaling violation effects.
At low $x_{\scr p}$ the difference is due to the mass of the hadrons, with a
cut-off at  $x_{\scr p} \simeq 2m_{\scr h}/E_{\scr CM}$.
 At high momentum, substantial scaling violation is expected because of running of $\alpha_{\scr s}$, 
as it is indeed observed for pions cross-sections. 
This behavior for pions is well reproduced by most of the fragmentation models, 
as shown for the case of JETSET~\cite{bib:JETSET} in the same fig.~\ref{fig:csvsxp}(left). 
A similar behavior is expected also for the others hadrons, 
but the data show little scaling violation for kaons, and almost no violation for protons. 
These results, even if preliminary, represent already 
very useful inputs for tuning the simulation of fragmentation processes down to an energy of 10 GeV.  

 \subsection{Test of QCD predictions in the MLLA model}
These data can also be used  to test some predictions of QCD in the 
Modified Leading Logarithm Approximation (MLLA)~\cite{bib:MLLA}. 
For this purpose, it is convenient to plot the differential cross section as a function of the 
variable $\xi = -{\mathrm{ln}}(x{\scr p})$. 
MLLA predicts that a slightly distorted gaussian function should
be able to fit the data over almost the full kinematic range. 
Fig.~\ref{fig:MLLA} show the distributions for 
$\pi, K, \eta ~{\mathrm{and}}~ p/\bar{p}$ with the results of the fit. 
The distorted Gaussian is able to describe the \babar\ 
data at the few percent level over the full measured momentum region with reasonably small skewness and
kurtosis values,  consistent with theoretical predictions.\\
The position of the peak $\xi^{\scr \star}$ of the $\xi$ distribution is predicted to decrease 
exponentially with increasing particle mass. 
The $\xi^{\scr \star}$ values reported in table~\ref{tab:xipeak}
show a clear disagreement with this prediction: the value for pions is quite different from the others,
 but those for protons and kaons are very similar one another, 
then inconsistent with a continued exponential decrease.
This qualitative behavior is not peculiar of the \babar\ operational energy region, but it has
been already observed at higher energies and it is particularly evident in measurements performed 
at $\sim 90$ GeV.\\
\begin{figure}[htb]
\begin{center}
\includegraphics[height=.3\textheight,width=.55\textwidth]{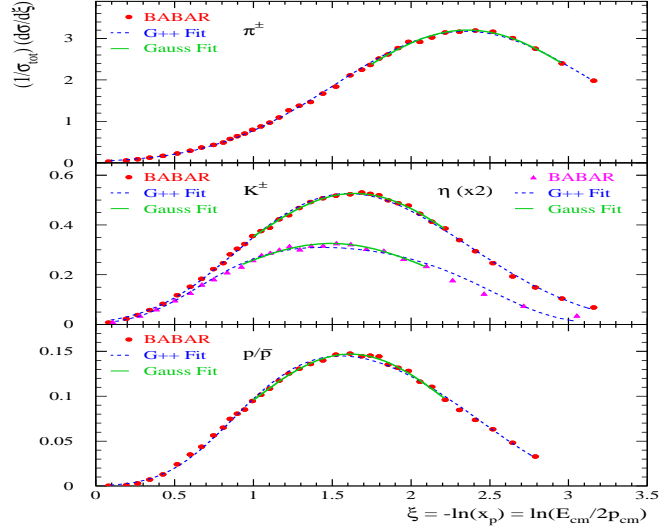}
\end{center}
\caption{Differential pion (top), kaon, eta (middle) and proton (bottom) 
 cross section in the continuum data (solid symbols) as a function of $\xi = -{\mathrm{ln}}(x{\scr p})$
shown with the result of a Gaussian fit to the data (solid lines) 
and of a distorted Gaussian ($G^{\scr ++}$, dashed lines).
The distorted gaussian is parametrized as: 
$G^{\scr ++}(\xi;N,\xi,\sigma,s,k) = \frac{N}{\sigma\sqrt{2\pi}} 
{\mathrm{exp}}\left( \frac{1}{8}k + \frac{1}{2}s\delta -
\frac{1}{4}(2+k)\delta^{\scr 2} + \frac{1}{6}s\delta^{\scr 3} 
+ \frac{1}{24}k\delta^{\scr 4} \right);~~$ 
where $\delta = (\xi - \bar{\xi})/\sigma$, $\bar{\xi}$ is the mean of the $\xi$ distribution, 
$\sigma$ is the square root of its variance, $s$ is the skewness and $k$ the kurtosis.  
}
\label{fig:MLLA}
\end{figure}

\begin{table}[htb]
\caption{Position of the peaks $\xi^{\scr \star}$ of the 
$\xi$ distributions for  $\pi, ~k, ~\eta ~{\mathrm{and}} ~p(\bar{p})$.}
\label{tab:xipeak}
\vspace{0.4cm}
\begin{center}
\begin{tabular}{|l|c|}
\hline
particle type &  $\xi^{\scr \star}$ \\
\hline
\pipiu\       &  $2.36 ~\pm ~0.01$ \\
\kpiu\        &  $1.64 ~\pm ~0.01$ \\
$\eta$        &  $1.44 ~\pm ~0.02$ \\
$p(\bar{p})$  &  $1.61 ~\pm ~0.01$ \\
\hline
\end{tabular}
\end{center}
\end{table}

\section{Conclusions}
Preliminary results from \babar\ on final states produced through ISR demonstrate the high physics 
potential of this sample, which should yield precise measurements of \epiu\emeno\ annihiliation
cross sections.
The ratio \ERRE, will be measured from the sum of exclusive channels, providing input for theoretical 
determination of the hadronic contribution to \gmenodue\ and \alphaqed. The preliminary \eetopioni\ 
cross section, from threshold up to 4.5 GeV  has been obtained, with a systematic error of about 5\% 
in the central region. 
The radiative return to \jpsi\ resonance allows to measure the relative branching 
fractions with best to date accuracy.\\
Preliminary measurements of the inclusive hadronic production cross sections have been performed 
at \babar\ with precision comparable to the previous measurements at $\sim 91$ GeV. 
These data will be very valuable for a better comprehension of the fragmentation process and in 
particular to study the scaling properties down to an energy of 10 GeV. 
The observed discre\-pancy between data and Monte Carlo predictions on scaling violation for etas, 
kaons and above all protons gives important experimental inputs for tuning models.\\

\end{document}
